\documentclass[letterpaper, 10 pt, conference]{ieeeconf} 
\IEEEoverridecommandlockouts
\usepackage{cite}
\usepackage{amsmath,amssymb,amsfonts}
\DeclareMathOperator*{\argmax}{arg\,max}
\DeclareMathOperator*{\argmin}{arg\,min}
\usepackage[ruled,vlined]{algorithm2e}
\usepackage{graphicx}
\usepackage{textcomp}
\usepackage{siunitx}
\usepackage{xfrac}
\usepackage{makecell}

\usepackage{xcolor}
\def\BibTeX{{\rm B\kern-.05em{\sc i\kern-.025em b}\kern-.08em
    T\kern-.1667em\lower.7ex\hbox{E}\kern-.125emX}}

\usepackage{hyperref}
\usepackage{booktabs}
\usepackage{multirow}
\let\labelindent\relax
\usepackage[inline]{enumitem}

\newcommand{\AR}[1]{\textcolor{cyan}{[AR: #1]}}

\begin{document}

\title{\LARGE \bf
Plasma Spray Process Parameters Configuration\\using Sample-efficient Batch Bayesian Optimization
}

\author{
Xavier Guidetti$^{1,2}$,
Alisa Rupenyan$^{1,2}$, 
Lutz Fassl$^{3}$,
Majid Nabavi$^{3}$, and
John Lygeros$^{1}$

\thanks{This project has been funded by the Swiss Innovation Agency (Innosuisse), grant Nr. 37896, and by the Swiss National Science Foundation under NCCR Automation.}
\thanks{$^1$ Automatic Control Laboratory, ETH Zürich, Switzerland}
\thanks{$^2$ Inspire AG, Zürich, Switzerland}
\thanks{$^3$ Equipment Digitalization Team, Oerlikon Metco, Switzerland \newline
\texttt{\{xaguidetti, ralisa, lygeros\}@control.ee.ethz.ch} \newline
\texttt{\{lutz.fassl, majid.nabavi\}@oerlikon.com}}
}

\maketitle

\thispagestyle{empty}
\pagestyle{empty}

\begin{abstract}
Recent work has shown constrained Bayesian optimization to be a powerful technique for the optimization of industrial processes. In complex manufacturing processes, the possibility to run extensive sequences of experiments with the goal of finding good process parameters is severely limited by the time required for quality evaluation of the produced parts. To accelerate the process parameter optimization, we introduce a parallel acquisition procedure tailored on the process characteristics. We further propose an algorithm that adapts to equipment status to improve run-to-run reproducibility. We validate our optimization method numerically and experimentally, and demonstrate
that it can efficiently find input parameters that produce the desired outcome and minimize the process cost.

\end{abstract}




\section{Introduction}

One of the main challenges in manufacturing is to quickly find a combination of input parameters that optimize the manufacturing process, in terms of productivity and resulting quality. This is especially true for processes that are difficult to model due to their dependence on multiple process inputs and having multiple correlated characteristics as an outcome. Atmospheric plasma spraying (APS) belongs to this class of processes: the number of trials to find an appropriate process window is restricted by the lengthy and expensive procedure to determine the resulting mechanical properties of the manufactured part and the requirements on the quality outcomes are set independently of their mutual correlations. We propose a sample-efficient data-driven approach to find optimal process parameters for a desired quality specification and we apply the method to an industrial APS process.

APS is a thermal spraying process where micrometer-sized powder particles are injected into a viscous enthalpy plasma jet that heats them and propels them. The particles form a protective coating that improves the mechanical properties of a substrate upon bonding with its surface. The coating properties (application rate, thickness, porosity, microhardness) depend on multiple process input parameters such as for example primary gas flow-rate, secondary gas flow-rate, gun current, carrier gas flow-rate, powder feed-rate, and stand-off distance \cite{Leblanc2000StudySpraying}. The parameters are usually selected and adapted by experienced operators and adjusted following an intensive trial and error procedure. As it is not feasible to model an APS process with a physics-based model due its variability and sensitivity in industrial conditions, the standard approach for process set-up and optimization is based on statistical principles. Candidate combinations of process parameters are proposed following a full factorial or fractional design of experiments, optimal orthogonal designs or Taguchi arrays to limit the number of experiments \cite{Gao_2012}. Then, the best candidates are selected following regression analysis of the experimentally determined quality parameters. Alternatively, the process parameters can be adjusted sequentially after the quality analysis that follows each run. Data-driven modeling using neural networks \cite{Wu2015EmpiricalProcess, Datta2013ModelingAnalysis} has been proposed to relate process inputs to quality parameters and mechanical or structural properties of the substrate, verifying experimentally the quality parameters on a run-to-run basis \cite{Kanta2008ArtificialProcesses, Kanta2009ArtificialAttributes}. As evaluations are expensive and time-consuming, the efficacy of both statistical and model-based process optimization techniques is compromised by the limited number of experiments included. Additionally, a restricted subset of the process input parameters is normally considered, which fails to explain the effects observed on larger data sets. Other approaches \cite{Friis2003ControlWindows, Sampath2009SensingCorrelations} propose to stabilize the process through the exploitation of real-time plasma characteristics sensors. Unfortunately, due to the required complex set-up for plasma sensing, these methods are not suited to commercial applications and ultimately do not offer a meaningful improvement over full input parameters modeling.

All discussed process optimization methods neglect the information made available during the set-up process, counting either on prior knowledge within the learned models, or on statistical insights from the collected data. Previous work on data-driven optimization of manufacturing processes \cite{Maier2019BayesianTurning, MaierSelf-OptimizingOptimization} has shown that Bayesian optimization (BO) is an information efficient and effective technique to automate the set-up of industrial processes in a limited amount of experiments.

In this paper we propose a BO-based approach to selecting and maintaining optimal input parameters with respect to the desired quality parameters of produced coatings. The main contributions of our work are: 
\begin{enumerate*}
    \item a novel candidate selection procedure, prioritizing feasible process parameters,
    \item an input-output hybrid modeling structure exploiting both data and physical knowledge, and
    \item a parallelized optimization procedure that adapts to the process' parametric drift.
\end{enumerate*}

In section \ref{sec_problem}, we detail the challenge that motivated the research. Section \ref{sec_background} describes the techniques upon which we base our work. Section \ref{sec_method}  introduces the parallelized BO-based method for process parameter optimization. 
Finally, sections \ref{sec_mod_res} and \ref{sec_opt_res} contain the results corresponding respectively to the proposed modeling and optimization methods.


\section{Problem Formulation} \label{sec_problem}

\begin{figure}[htbp]
\centerline{\includegraphics{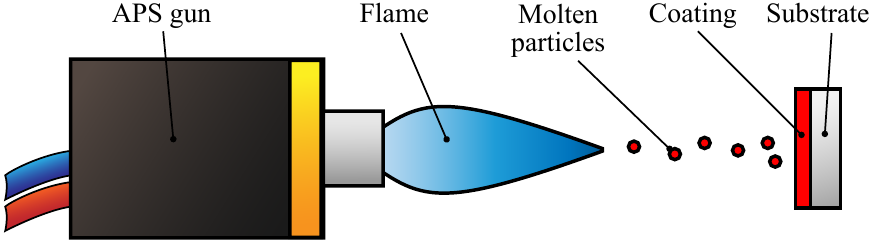}}
\caption{Schematics of the APS process}
\label{fig:process}
\end{figure}

The typical APS process depends on multiple interconnected parameters that control the physics of the process. A schematic representation is shown in Fig. \ref{fig:process}, which outlines the main components of a typical APS set-up. Our goal is to select values for six controllable inputs to regulate coating properties such as application rate, deposition efficiency, microhardness, and porosity and minimize the process cost. The powder type depends on the application and is kept fixed during configuration. While spraying a coating, we measure the gun voltage: it contains valuable information about the process -- namely, the equipment status. The most challenging and practically useful goal in APS is to discover input parameters that maintain microhardness and porosity within predefined limits, while maximizing the lifetime of the electrode and nozzle in the APS gun. In the absence of real-time measurements reflecting the state of the system, the objective of maximizing the lifetime is encoded through the minimization of the stress index, an empirical relation reflecting the working conditions of the gun components during the coating process. The stress index value can be calculated explicitly, given the input parameters.

Each evaluation of input parameters in APS corresponds to lengthy and expensive coating and measurement procedures. The main focus of the optimization lies in finding \emph{feasible} samples, while cost reduction (minimization of the stress index) is a secondary goal to be achieved once feasibility has been reached. Furthermore, as this process is also used to manufacture very small lots, producing as many feasible samples as possible during configuration itself is important \cite{I4_0review}. While suboptimal in terms of cost, these samples are usable, reducing the amount of runs conducted and the total configuration cost. Finally, process variation is unavoidable, and the coating process often suffers from undocumented changes in the corresponding setup, drifts, and variations in the used coating powders. Parallelizing the exploration phase, prioritizing feasibility, and incorporating available data from process monitoring are required for sample-efficient process parameter configuration.

\section{Background} \label{sec_background}

\subsection{Gaussian processes}

We model each constraint function $c(\mathbf{x})$ using Gaussian process regression. A Gaussian process (GP) is a collection of random variables, any finite number of which have a joint Gaussian distribution. It provides a distribution over functions $c(\mathbf{x}) \sim \mathcal{GP}(\mu(\mathbf{x}),k(\mathbf{x},\mathbf{x}'))$ that is fully defined by its mean function $\mu(\mathbf{x})$ and its covariance, given by the kernel function $k(\mathbf{x},\mathbf{x}')$. We denote the $i$-th measurement corresponding to an input vector $\mathbf{x}_i$ by $y_i=c(\mathbf{x}_i)+\varepsilon_i$, where $\varepsilon_i$ is the measurement noise with distribution $\mathcal{N}(0,\sigma^2_\mathrm{n})$. Given a set of $p$ input vectors paired with the corresponding noise corrupted measurements $\mathcal{T} = \{(\mathbf{x}_i,y_i)\}_{i=1}^p$, we can calculate the posterior distribution of $c(\cdot)$ at any query point $\bar{\mathbf{x}}$. Denoting the set of inputs $\mathbf{X} = \{\mathbf{x}_i\}_{i=1}^p$ and the set of corresponding measurements $\mathbf{y} = \{y_i\}_{i=1}^p$, we obtain $\Tilde{c}(\bar{\mathbf{x}}) \sim \mathcal{N}\left(\mu_c(\bar{\mathbf{x}}),\sigma^2_c(\bar{\mathbf{x}})\right)$, where the corresponding posterior mean and variance are given as
\begin{align}
\mu_c(\bar{\mathbf{x}}) &= \mu(\bar{\mathbf{x}}) + k(\bar{\mathbf{x}},\mathbf{X})[k(\mathbf{X},\mathbf{X})+\sigma^2_\mathrm{n}\mathbb{I}]^{-1}(\mathbf{y}-\mu(\mathbf{X})) \, , \nonumber \\
\sigma^2_c(\bar{\mathbf{x}}) &= k(\bar{\mathbf{x}},\bar{\mathbf{x}}) - k(\bar{\mathbf{x}},\mathbf{X})[k(\mathbf{X},\mathbf{X})+\sigma^2_\mathrm{n}\mathbb{I}]^{-1}k(\mathbf{X},\bar{\mathbf{x}}) \nonumber \, .
\end{align}

After model training, the posterior mean $\mu_c(\bar{\mathbf{x}})$ and variance $\sigma^2_c(\bar{\mathbf{x}})$ can be used respectively as the model prediction and corresponding uncertainty at point $\bar{\mathbf{x}}$. To do so, a confidence interval of 95\% is typically selected.
A complete overview of GPs and their practical use can be found in \cite{Rasmussen2006GaussianLearning}.

\subsection{Parallel Constrained Bayesian Optimization} \label{sec_parallelCBO}

In its simplest form, BO is a sequential strategy for the optimization of expensive-to-evaluate functions. The method starts by placing a prior over the unknown objective function and then updates it with the collected data to form a posterior distribution of the objective. BO is commonly used with GP models, which use the available evaluations to produce a probabilistic distribution of the objective and can be updated when new samples are added to the known experiment set. 
The posterior distribution is used by an acquisition function to select the next candidate for evaluation, trading off exploration and exploitation by combining the information content at the inputs and the corresponding predicted performance. Optimization constraints can be modeled as separate GPs \cite{Gardner2014BayesianConstraints} and the inputs corresponding to fulfilling the constraints with high probability are chosen by adapting the acquisition function. The approach has been used on manufacturing processes \cite{Maier2019BayesianTurning, Maier_2020}, for the tuning of cascade controllers \cite{konig2020safety, Khosravi2021}, and for improving precise position tracking \cite{rupenyan2021performance, konig2021safe}. Another approach to respect safety constraints, known as safe BO (SafeOpt), provides probabilistic guarantees that all candidate samples remain in the constraint set \cite{Sui} and has been demonstrated for robotic applications \cite{Berkenkamp2} and for adaptive control in position tracking \cite{konig2021safe}.

Due to the complexity of the measurement process, our approach also requires the generation of batches of query points that will be evaluated simultaneously. Several methods to parallelize BO have been proposed \cite{Snoek2012PracticalAlgorithms, Wang2020ParallelFunctions, Ginsbourger2008AProcesses, Azimi2012HybridOptimization}. We utilize a simple fixed batch size technique that is largely based on sequential selection of query points. Unlike standard BO where the candidate $\mathbf{x}^*$ evaluation is conducted immediately, a prediction $\hat{y}^*$ of the output produced by the candidate is made. The data set of known evaluations is virtually expanded using the prediction and another candidate is selected. The process repeats until a batch having the desired size $n$ has been generated. The points belonging to the batch are evaluated simultaneously and the predictions $\{\hat{y}^*_i\}_{i=1}^n$ are replaced with the experimental results $\{y^*_i\}_{i=1}^n$. When using GPs to model the unknown functions, data set virtual augmentation can be carried on by drawing samples $\hat{y}^*$ from the posterior distribution evaluated at the candidate site $\Tilde{c}(\mathbf{x}^*) \sim \mathcal{N}\left(\mu_c(\mathbf{x}^*),\sigma^2_c(\mathbf{x}^*)\right)$.

\section{Method} \label{sec_method}

Industrial process modeling often includes uncontrollable but measurable parameters that correlate with the current status of manufacturing equipment, such as gun voltage. APS quality parameters collected in different sessions usually contain offsets due to drifts or undocumented changes to the equipment between sessions. Including voltage in the modeling makes it possible to compensate for these deviations.
The proposed formulation also contains parameters that are fixed according to a design choice, such as the powder type. Depending on the application, the powder selected prior to any optimization and remains unchanged. It is excluded from the pool of optimization variables, but necessary for accurate modeling.

The controllable inputs combinations $\mathbf{x}_\mathrm{c}$ which affect the coating process belong to a bounded domain $\mathcal{X}_\mathrm{c} \subset \mathbb{R}^6$ according to process limitations. We further expand $\mathbf{x}_\mathrm{c}$ with the desired powder and the measured gun voltage collected in $\mathbf{x}_\mathrm{m}$ to define the input vector $\mathbf{x} = (\mathbf{x}_\mathrm{c},\mathbf{x}_\mathrm{m})$, which contains all the modeled parameters that generate a coating with specifications $\mathbf{c}(\mathbf{x})$.

We write our optimization problem as
\begin{align}
\min_{\mathbf{x}_\mathrm{c} \in \mathcal{X}_\mathrm{c}}\quad &S(\mathbf{x}) \label{eq:opt_problem} \\ 
\text{s.t} \quad\; & \lambda_1^L \leq c_1(\mathbf{x}) \leq \lambda_1^H \ , \nonumber
& \lambda_2^L \leq c_2(\mathbf{x}) \leq \lambda_2^H \ , \nonumber
\end{align}
where $S(\mathbf{x})$ denotes the process cost and $c_1(\mathbf{x})$ and $c_2(\mathbf{x})$ the bounded coating properties. $S(\mathbf{x})$ can be freely computed as a deterministic function of a subset of the inputs in $\mathbf{x}_\mathrm{c}$, but the explicit formula is omitted for reasons of confidentiality. $c_1(\mathbf{x})$ and $c_2(\mathbf{x})$ are unknown and measured from the process. Their queries are conducted in batches $\mathcal{B} = \{\mathbf{x}_i|{\mathbf{x}_\mathrm{c}}_i \in \mathcal{X}_\mathrm{c}\}^n_{i=1}$. We assume that the optimization can be initialized with a set of previously evaluated input vectors and corresponding constraints values $\mathcal{T}=\{\mathbf{x},\mathbf{c}(\mathbf{x})\}$.


\subsection{Acquisition Procedure}

We propose a custom-build acquisition procedure that has been tailored to our application. We first introduce two functions: \textit{improvement} and \textit{feasibility probability}. The first one determines the amount of improvement, in terms of cost reduction, that a vector with a candidate combination of inputs can produce. We define it as
\begin{equation}
    I(\mathbf{x}) = \max \left\{0, S(\mathbf{x}^+)-S(\mathbf{x}) \right\} \label{eq:imp}\\,
\end{equation}
where $\mathbf{x}^+$ is the feasible combination of inputs with the lowest cost found so far. If no feasible point is known, we set $S(\mathbf{x}^+) = \max_{\mathbf{x}_\mathrm{c} \in \mathcal{X}_\mathrm{c}}S(\mathbf{x})+1$. Candidate combinations producing a cost higher than $\mathbf{x}^+$ return no improvement. In contrast with \cite{Gardner2014BayesianConstraints}, we do not need to take an expectation of \eqref{eq:imp} as $S(\mathbf{x})$ is deterministic.

To include the constraints of \eqref{eq:opt_problem} we define the feasibility probability as
\begin{align}
    \text{\em{FP}}(\mathbf{x}) &= \text{\em{Pr}}[\lambda^L \leq \Tilde{c}(\mathbf{x}) \leq \lambda^H] \\ \nonumber
    &= \int_{\lambda^L}^{\lambda^H} p(\Tilde{c}(\mathbf{x})|\mathbf{x},\mathcal{T})d\Tilde{c}(\mathbf{x}) \ .
\end{align}
Since $\Tilde{c}(\mathbf{x})$ has Gaussian marginals, $\text{\em{FP}}(\mathbf{x})$ is a Gaussian cumulative distribution function. Following our experiments, we can assume constraints independence. Then, the feasibility probability is given by
\begin{equation}
    \text{\em{FP}}(\mathbf{x}) = \prod_{k=1}^K \text{\em{Pr}}[\lambda_k^L \leq \Tilde{c}_k(\mathbf{x}) \leq \lambda_k^H] \label{eq:fp}\ .
\end{equation}

We now define the two acquisition functions that our algorithm exploits: 
\begin{subequations}
\begin{align}
    \alpha_{\text{\em{FIP}}}(\mathbf{x}) &= \text{\em{FP}}(\mathbf{x})\text{sgn}\{I(\mathbf{x})\} \label{fip} \\
    \alpha_{\text{\em{HFI}}}(\mathbf{x}) &= (\text{\em{FP}}(\mathbf{x})-\pi)I(\mathbf{x}) \label{hfi} \ ,
\end{align}
\end{subequations}
where $\pi \in [0, 1]$ is a confidence threshold that tunes the aggressiveness of our acquisition algorithm. \textit{Feasible Improvement Probability} (FIP) \eqref{fip} returns conservatively the feasibility probability of the candidates that are known to produce any cost improvement. As the magnitude of the improvement is eliminated by the sign function, maximizing $\alpha_{\text{\em{FIP}}}(\mathbf{x})$ corresponds to looking for the points that have the highest chances of respecting the constraints. \textit{High FIP Improvement} (HFI) \eqref{hfi} is more aggressive: the magnitude of the cost improvement modulates the candidate selection, producing a trade-off between feasibility probability and reward. Maximizing $\alpha_{\text{\em{HFI}}}(\mathbf{x})$ returns candidates that reduce remarkably the cost while maintaining a minimum feasibility probability of $\pi$. \par

\begin{algorithm}[ht]
\SetAlgoLined
\LinesNumbered
\SetKwInOut{Input}{input}
\Input{$\text{\em{FP}}(\mathbf{x})$ and $I(\mathbf{x})$ of all candidates $\mathbf{x}$ in the candidates set $\mathcal{U}$, previously evaluated inputs set $\mathcal{T}$, constraints $\lambda$, threshold probability $\pi$}
Compute $\alpha_{\text{\em{FIP}}}(\mathbf{x})$ and $\alpha_{\text{\em{HFI}}}(\mathbf{x})$ for all candidates\;
Group the feasible elements of $\mathcal{T}$ in $\mathcal{T}_\mathrm{f} \subset \mathcal{T}$\;
\eIf{$\mathcal{T}_\mathrm{f} = \emptyset$}{
$\alpha(\mathbf{x}) \longleftarrow \alpha_{\text{\em{FIP}}}(\mathbf{x})$ for all $\mathbf{x} \in \mathcal{U}$ \label{line_nofeas}\;
}{
\eIf{any candidate verifies $\alpha_{\text{\em{FIP}}}(\mathbf{x}) > \pi$ \label{line_pi}}{
$\alpha(\mathbf{x}) \longleftarrow \alpha_{\text{\em{HFI}}}(\mathbf{x})$ for all $\mathbf{x} \in \mathcal{U}$\;}{
$\alpha(\mathbf{x}) \longleftarrow \alpha_{\text{\em{FIP}}}(\mathbf{x})$ for all $\mathbf{x} \in \mathcal{U}$ \label{line_low_feas}\;}
}
\Return{selected candidate $\mathbf{x}^* = \argmax_{\mathbf{x} \in \mathcal{U}} \alpha(\mathbf{x})$ \label{line_maxim}}
 \caption{Candidate Selection}
 \label{alg_candsel}
\end{algorithm}


Given any set of possible candidates $\mathcal{U}$, algorithm \ref{alg_candsel} presents the complete candidate selection procedure. In preliminary experiments, we observed that acquisition procedures such as the one presented in \cite{Gardner2014BayesianConstraints} were too aggressive, while an approach focusing on FIP only led to excessively conservative exploration. Given that maintaining feasible solutions has higher priority than optimizing $S(\mathbf{x})$, we formulate an acquisition procedure that maintains a trade-off between the two approaches. As long as no feasible point is found, we exclusively focus on maximizing the chances of finding one. We therefore perform the optimization according to \eqref{fip} (line \ref{line_nofeas}). Once we have found a feasible experiment, we take a mixed approach. We want to ensure that the aggressive exploration of \eqref{hfi} is conducted only with a sufficient safety margin, given by the confidence threshold $\pi$ (line \ref{line_pi}). When the probability of finding cost reducing points is too low, we take the conservative approach of \eqref{fip} (line \ref{line_low_feas}). At line \ref{line_maxim}, we select the candidate belonging to the candidates set $\mathcal{U}$ that has been assigned the largest $\alpha(\mathbf{x})$. This procedure increases the amount of feasible experiments found during the optimization and favors a safer exploration of the constraints space over uncertain large improvements. Given the deterministic nature of our objective function, both \eqref{fip} and \eqref{hfi} consider only candidates that will certainly produce a cost reduction.

\subsection{Quality Specifications Modeling} \label{sec_model_structure}

All four coating properties are modeled independently, with single-output GPs. Eight inputs are required: the six controllable inputs, the powder type, and the measured gun voltage. The GP models use exact inference and a squared exponential automatic relevance determination (ARD) kernel. The models for application rate, porosity, and deposition efficiency are initialized with equal hyper-parameters and a zero mean function. We then find suitable hyper-parameters by maximization of the marginal likelihood on the training data.
Microhardness modeling can be improved by the addition of empirical knowledge to the GP model structure. Previous works \cite{Thirumalaikumarasamy2015EffectAlloy, Kanta2008ArtificialProcesses, Chen2004InfluenceCoatings} highlight a positive correlation between the gun absorbed power and coating's microhardness, for different materials coated using APS. Our sensitivity analysis conducted on the collected data indicates that the process inputs having the highest effect on the absorbed power are secondary gas flow-rate (correlating negatively) and measured gun voltage (correlating positively). To include this effect, we incorporate in the GP model a linear mean function linking each of the two relevant inputs to microhardness. We constrain the secondary gas flow-rate mean function coefficient to be negative and the voltage mean function coefficient to be positive. We initialize both at zero and then maximize the marginal likelihood with respect to all the hyper-parameters simultaneously. The described mean function can be written as
$\mu(\mathbf{x}) = \sum_{i=1}^{8}\gamma_i x_i$, where $\gamma_i < 0$ for the secondary gas flow-rate, $\gamma_i > 0$ for the voltage and $\gamma_i = 0$ for all other $x_i$.
Incorporating this coarse empirical knowledge improves our predictions when an unexpected offset in the measured voltage requires the model to extrapolate from well-modeled areas, at the expense of only two additional hyperparameters. 
Given the small size of our data set, the efficiency of BO and the run-to-run application, the resulting four models are easily tractable despite the computational complexity of GPs.

\subsection{Optimization Workflow}



To prevent offsets induced by variations in the equipment status, we incorporate in our models process monitoring data, such as gun voltage measurements. In the input vector $\mathbf{x} = (\mathbf{x}_{\mathrm{c}},\mathbf{x}_{\mathrm{m}})$, only the parameters in $\mathbf{x}_{\mathrm{c}}$ can be freely tuned during BO. The voltage contained in $\mathbf{x}_{\mathrm{m}}$ depends on the controllable parameters as well as on the current equipment status. We assume that this status remains constant during a single experimental session but can change between sessions. At the beginning of each batch, we conduct a short experiment to estimate the current equipment status from the voltage in $\mathbf{x}_{\mathrm{m}}$. We then adapt the controllable parameters in $\mathbf{x}_{\mathrm{c}}$ to our estimate when proceeding with BO.

Using the entries of the initialization data set $\mathcal{T}$, we train a model accepting the controllable inputs $\mathbf{x}_\mathrm{c}$ and predicting the gun voltage $\hat{V} = \mathrm{M_V} (\mathbf{x}_\mathrm{c})$. At the beginning of an experimental session, we ignite the gun with any settings $\mathbf{x}_\mathrm{c}^\mathrm{b} \in \mathcal{T}$, measure the corresponding voltage $V^\mathrm{b}$ and compute the voltage offset $\delta^\mathrm{b} = V^\mathrm{b} - \hat{V}^\mathrm{b}$, where $\hat{V}^\mathrm{b} = \mathrm{M_V} (\mathbf{x}_\mathrm{c}^\mathrm{b})$ is the measurement that an unchanged equipment -- with respect to $\mathcal{T}$ -- would produce. 

To generate the list of candidates that will be evaluated by the BO procedure, we first grid the space of controllable inputs $\mathcal{X}_\mathrm{c}$ to produce a set of 20000 candidates $\mathbf{x}_\mathrm{c}$. 
We then predict the voltage $\hat{V}_\delta$ corresponding to each candidate $\mathbf{x}_\mathrm{c}$ in the set according to $\hat{V}_\delta = \mathrm{M_T}(\mathbf{x}_\mathrm{c})+\delta^\mathrm{b}$. Each voltage prediction and the (fixed) desired powder, collected in $\mathbf{x}_\mathrm{m}$, are used to expand the corresponding $\mathbf{x}_\mathrm{c}$, producing a set $\mathcal{U}$ of vectors $\mathbf{x} = (\mathbf{x}_\mathrm{c},\mathbf{x}_\mathrm{m})$.



\begin{algorithm}[htbp]
\SetAlgoLined
\LinesNumbered
\SetKwInOut{Input}{input}
\Input{Initialization data set $\mathcal{T}$, candidates set $\mathcal{U}$, constraints $\lambda$, batch size $n$, threshold probability $\pi$, termination threshold $\epsilon$}
Create an empty candidates batch $\mathcal{B} \leftarrow \emptyset$\;
\Repeat{the termination condition is met\label{alg_line_termination}}{
\If{$\mathcal{B} \neq \emptyset$}{
Evaluate experimentally the candidates in $\mathcal{B}$ and expand $\mathcal{T}$ with $\{(\mathbf{x}_i^*,\mathbf{y}_i^*)\}_{i=1}^n$, where $\mathbf{y}_i^*$ collects the evaluations of each $\mathbf{x}_i^*$\;
Empty $\mathcal{B} \leftarrow \emptyset$\;
} 
Use \eqref{eq:imp} to calculate $I(\mathbf{x})$ of all $\mathbf{x} \in \mathcal{U}$, make a virtual copy $\mathcal{T}_\mathrm{v} \leftarrow \mathcal{T}$\;
    \For{$i=1$ \KwTo $n$}{
    Using the the data in $\mathcal{T}_\mathrm{v}$, model the constraints $\mathbf{c}(\cdot)$ according to section \ref{sec_model_structure}\;
    Use \eqref{eq:fp} to calculate $\text{\em{FP}}(\mathbf{x})$ of all $\mathbf{x} \in \mathcal{U}$\;
    Select the candidate $\mathbf{x}^*_i$ using Alg. \ref{alg_candsel}\;
    Remove $\mathbf{x}^*_i$ from $\mathcal{U}$ and add it to $\mathcal{B}$\;
    Expand $\mathcal{T}_\mathrm{v}$ with $(\mathbf{x}^*_i,\mu_\mathbf{c}(\mathbf{x}^*_i))$, where $\mu_\mathbf{c}(\mathbf{x}^*_i)$ collects the predictive means of the constraints $\mathbf{c}(\cdot)$ \label{alg_line_expand}\;
    $i \leftarrow i+1$\;
    }
}

\Return{feasible stress index minimizer $\mathbf{x}^+ = \argmin_{\mathbf{x} \in \mathcal{T}_\mathrm{f}} S(\mathbf{x})$, where $\mathcal{T}_\mathrm{f} \subset \mathcal{T}$ contains the feasible elements of $\mathcal{T}$}
\caption{Parallel Optimization}
\label{alg_workflow}
\end{algorithm}

\begin{figure}[htbp]
\centerline{\includegraphics{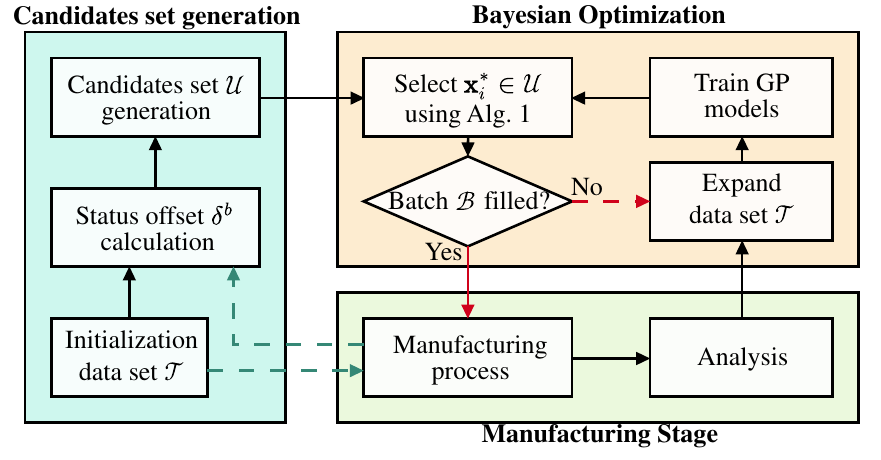}}
\caption{Summary of the proposed configuration method}
\label{fig:summary}
\end{figure}

Parallel BO is carried on on the candidates set $\mathcal{U}$ as detailed in Algorithm \ref{alg_workflow}. Following the procedure described in section \ref{sec_parallelCBO}, we select candidates individually and expand our virtual database using the GPs posterior mean (line \ref{alg_line_expand}). The termination condition required at line \ref{alg_line_termination} is met when at least half of the candidates $\mathbf{x}^*_i$ selected for a batch $\mathcal{B}$ verify $\alpha_{\text{\em{FIP}}}(\mathbf{x}^*_i) < \epsilon$. This interrupts the procedure when most of the candidates have a low FIP, indicating that further improvement is unlikely.

\section{Modeling Results} \label{sec_mod_res}

\subsection{Training Data Collection} \label{data_coll}

Coating trials were conducted on an Oerlikon Metco F4MB-XL gun, connected to a MultiCoat controller. The system consists of a rotating sample support paired with an APS gun mounted on a robot arm. 
Based on a fractional factorial design, a training data set of 86 experiments was collected. The continuous inputs space was discretized by setting the six controllable input parameters at two, three or four different levels according to their influence on the results. Two additional levels came from testing two different powders. The carefully chosen experimental design allowed to cover the feature space effectively while keeping the data collection cost reasonable. For each experiment, the four quality outputs were measured by the Oerlikon Metco metallurgical analysis laboratory. Additionally, we recorded the measured gun voltage. 
To quantify the effect of the combined process and measurement noises on the outputs, the collected data set contains 13 baseline runs, which have been conducted with identical controllable input parameters and powder. 

\subsection{Modeling Validation}

For each coating property, a GP regression was conducted as described in section \ref{sec_model_structure}. The models were trained using the data described in section \ref{data_coll}. We then generated a validation data set in experiments that produced 23 coated samples, including eight baseline runs with identical input parameters. To reduce experimental cost, we used only one of the two powders that were studied in the training data set. This produced coatings belonging to the low porosity and high microhardness region. Seven experiments were selected following domain expertise as particularly challenging combinations of inputs, covering also combinations outside of the training ranges. We further selected five combinations with both inputs and outputs sufficiently distant from the training data set entries and having the lowest GP posterior variance in the predicted coating properties. Four entries were generated as a first batch of the optimization procedure and used for model validation too.

\begin{figure}[htbp]
\centerline{\includegraphics{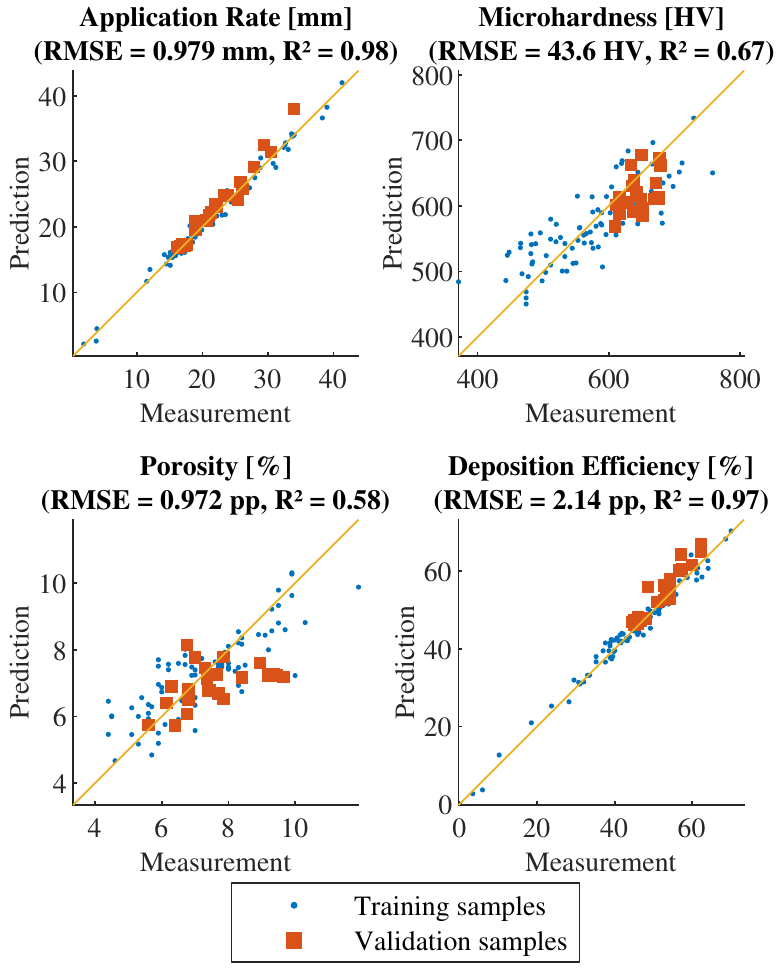}}
\caption{Comparison between predictions and measurements for the four quality specifications of training and validation samples.} 
\label{fig_exp_val}
\end{figure}

Fig. \ref{fig_exp_val} shows the error metrics, root-mean-square error (RMSE) and coefficient of determination (R$^2$), on the aggregate data set obtained by merging training and validation data. The choice is motivated by the observation that (with the exception of application rate) the validation measurements of quality parameters belong to a narrow range whose size is comparable with the process and measurement noise. While having little effect on the root-mean-square error, this would affect negatively the calculation of the coefficient of determination, making it insignificant. Application rate and deposition efficiency are predicted with remarkable accuracy, as shown both by the low RMSE and the R$^2$ close to $1$. Microhardness and porosity clearly suffer from the large process and measurement uncertainties that limit the prediction accuracy. The uncontrollable voltage offset impacting microhardness, requires the models to make predictions for uncharted areas of the inputs space, where uncertainty is larger. Nevertheless, the RMSE values and predictions dispersion are compatible with the standard deviations of the baseline experiments present in the training data set. These amount to \SI{44.7}{HV} for microhardness and \SI{0.891}{pp} (percentage points) for porosity, indicating that the models performance corresponds to the data quality. The lower $R^2$ values can be attributed to the low signal-to-noise ratio that the process and measurement uncertainties cause.

The models have higher performance when creating an aggregate data set (merging training and validation data) and then splitting it randomly for cross-validation. This is however a simpler task, given the nature of our problem. The main challenge is given by the unknown and uncontrollable change in equipment status between different experimental sessions. Training on a data set collected entirely in one session and validating on the data belonging to another session, as presented in this work, is the hardest task to address, and also the most realistic scenario that happens in practice.

\begin{figure}[htbp]
\centerline{\includegraphics{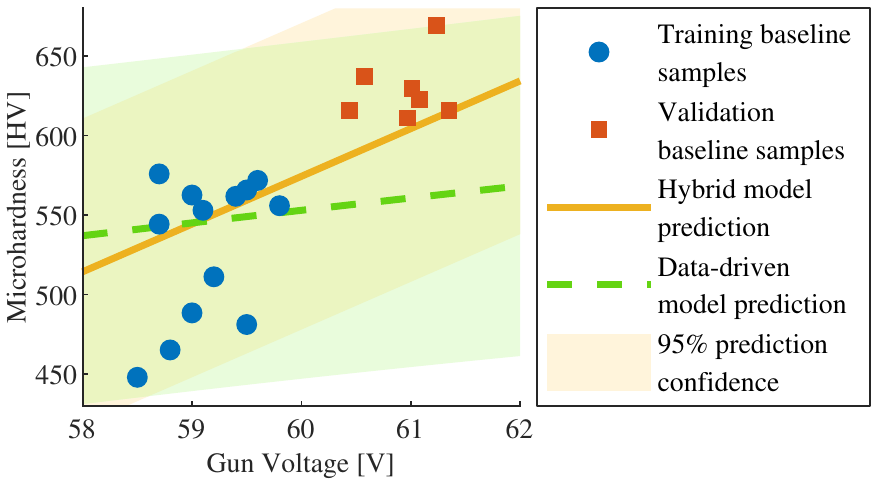}}
\caption{Comparison of microhardness predictions for the baseline samples, with and without taking into account the empirically known effect of gun power on microhardness.}
\label{fig_mhard_drift}
\end{figure}

In Fig. \ref{fig_mhard_drift} we highlight the effect of the voltage offset on the samples microhardness. We compare the measurements that were obtained for the 13 baseline runs coated in the training data set with the 7 baseline runs that were coated in the validation experiment and show model predictions. All the plotted samples have been coated with the same controllable inputs. The change in the measured voltage depends on a change in the equipment status between the two data collection sessions, and corresponds directly to a change in the gun power. These results show how the hybrid microhardness model, incorporating coarse empirical knowledge (cf. section \ref{sec_model_structure}), outperforms the purely data-driven model that is trained with no mean function.


\section{Optimization Results} \label{sec_opt_res}

\subsection{Simulated Process Optimization}\label{sec_simulated_res}


We first study numerically the performance of the proposed approach. For this, we build a neural network model using the training data set described in section \ref{data_coll}. It simulates the behavior of the APS machine and acts as an oracle during the optimization process, returning $c_1(\mathbf{x})$ and $c_2(\mathbf{x})$, the microhardness and porosity of virtual coated samples. It accepts eight inputs, the components of $\mathbf{x}$, and has one hidden layer of seven neurons. We add zero-mean Gaussian noise to the neural network outputs to replicate the uncertainty caused by the measurement procedure. Here, we choose to neglect the process noise. The measurement noise standard deviations have been estimated from repeated measurements of a set of samples coated simultaneously and amount to \SI{8.45}{HV} for microhardness and \SI{0.54}{pp} for porosity. We simulate a scenario where, during a first gun ignition, we measured a voltage offset of \SI{2}{V} from our expectation, indicating a change in the equipment status. We study only one type of powder and conduct five parallel experiments per batch. 

\begin{figure}[htbp]
\centerline{\includegraphics{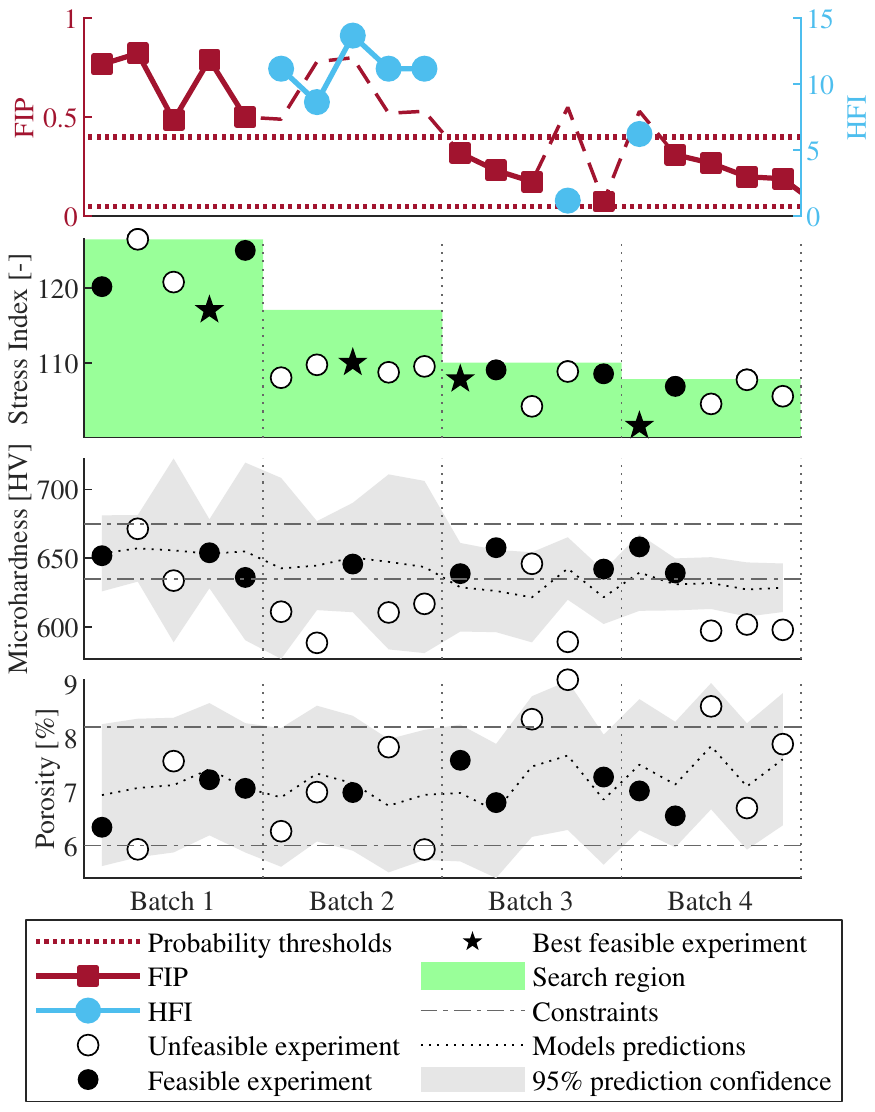}}
\caption{Simulated optimization of an APS process initialized with 86 experiments. The top panel shows the two acquisition functions (HFI is shown only when active). The next panel shows the minimization of the stress index, along with the best feasible point found so far. The bottom two panels show the evolution of the constrained microhardness and porosity for each subsequent batch, indicating the feasibility of each point and the associated prediction confidence interval.}
\label{fig:detailed_opt}
\end{figure}

The goal of the optimization is to find combinations of inputs producing coatings with microhardness in the range \SI{635}{HV} -  \SI{675}{HV}, and porosity between 6\% and 8.2\%, while minimizing the gun stress index. We select a batch size $n=5$, a threshold probability $\pi = 0.4$ and a termination threshold $\epsilon = 0.05$. Fig. \ref{fig:detailed_opt} shows in detail the optimization procedure for a case where the optimization algorithm has been initialized with $N_{\mathrm{init}}=86$ experiments, none of which respects the constraints. The initialization data set comprises the outputs of the simulated process, when fed with the inputs that generated the data set described in section \ref{data_coll}.  Initially, as no feasible point is known, the algorithm relies on the FIP acquisition function to maximize the chances of producing coatings respecting the constraints. The first batch consists of the five query points having the highest feasibility probability. The neural network oracle simulates the coating process and returns the corresponding coating properties of the chosen candidates. The first, fourth and fifth batch points turn out to be feasible, with the fourth point having the lowest cost. Its value is used as an upper bound of the search region in the subsequent batch.
The information from the five simulated experiments is used to expand the data set of known results and to retrace the GP models before the selection of the second batch is conducted. From this point on, as feasible candidates are known, HFI will be used as the acquisition function when FIP is above the activation threshold $\pi$, following Algorithm \ref{alg_candsel}. The optimization procedure is terminated after the fourth batch, when the values of FIP fall below the termination threshold $\epsilon$ for more than three candidates. 

Fig. \ref{fig:detailed_opt} shows that HFI is responsible for large and aggressive cost reductions in situations where the feasibility probability is high, while FIP takes a conservative approach that produces small cost improvements when the uncertainty about feasibility is significant. For microhardness, the GP predictive confidence is initially quite low, as we are targeting a region that is poorly modeled, because of the voltage offset. As the experiment progresses, the uncertainty is reduced to a size that is comparable to the standard deviation of Gaussian noise applied on the neural network outputs. For porosity, the predictive confidence bounds correspond to the standard deviation of the Gaussian noise already from the very beginning of the experiment and do not reduce further.

\begin{table}[htbp]
\caption{Minimum Cost and Stopping Batch for Different Initialization and Batch Sizes}
\label{tab1}
\centering
\renewcommand{\arraystretch}{1}
\begin{tabular}[h]{@{}l r r r c r r r @{}}\toprule
& \multicolumn{3}{c} {Batch size 5}  && \multicolumn{3}{c}{Batch size 10}\\
\cmidrule{2-4} \cmidrule{6-8}
$N_{\mathrm{init}}$ &  10 & 40 & 86  & &  10 & 40 & 86 \\
 \midrule
Cost   & 104 & 104 & 102 & & 110 & 104 & 102\\
Stopping batch & 9 & 6 & 4 && 4 & 3 & 3\\
Nr. evaluations & 45 & 30 & 20 & & 40 & 30 & 30 \\

\bottomrule
\label{tab_other_results}
\end{tabular}
\end{table} 

Table \ref{tab_other_results} shows the minimum cost found for different combinations of initialization data set and batch size, while maintaining $\pi = 0.4$ and $\epsilon = 0.05$ . We indicate the number of the batch corresponding to the termination criterion as well. As a point of comparison, the minimum feasible stress index found through grid search optimization of the noiseless neural network oracle was 98.8.

\subsection{Process Optimization Experiments}

\begin{figure}[htbp]
\centerline{\includegraphics{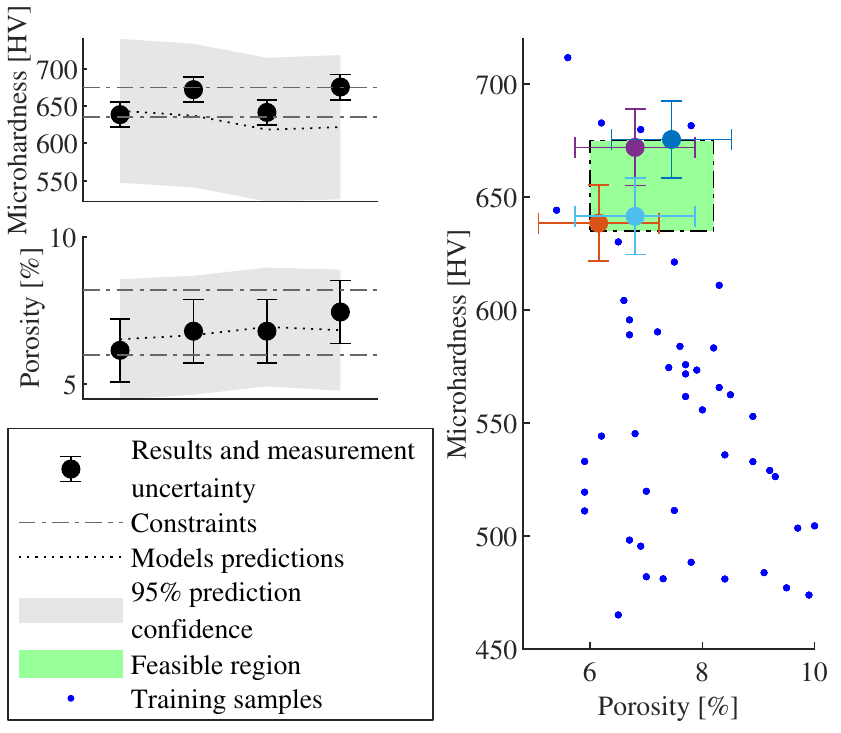}}
\caption{Experiments results for the optimization of an APS process showing the quality specifications of the four samples corresponding to the first optimization batch.}
\label{fig_experimental_opt}
\end{figure}

We now present a preliminary real-world validation of our method on the APS machine, where we conducted the first batch of experiments. 
The goal of the experiments was to find combinations of inputs that produce coatings with microhardness ranging between \SI{635}{HV} and \SI{675}{HV} and porosity between 6\% and 8.2\%. The ranges correspond to the numerical study, for the same type of powder and the same voltage offset of \SI{2}{V}. 
The GP models that are used by the optimization algorithm were initialized using all the available data set of 86 experiments. As in the case of the numerical APS optimization, there was no feasible experiment in the initialization data set, which forced the optimization algorithm to first look for the candidates having the highest feasibility probability. For its first batch, the algorithm returned five different combinations of inputs, of which two had very similar values. 
We therefore executed four experiments on the machine, resulting in the quality outputs shown in Fig. \ref{fig_experimental_opt}.

As discussed in section \ref{sec_simulated_res}, the voltage offset at each new experimental session causes the GP to have a large uncertainty, affecting mostly microhardness. 
The coated samples respect the imposed constraints both for microhardness and porosity. We do not plot the samples stress index as it is not optimized in the first (and only) batch which has been explored. The lowest found stress index belongs to the third sample and corresponds to 120.3. This value would be used as an upper bound in the second batch search, if the optimization needs to continue. If time constraints make further exploration undesirable or impossible, as in this experimental session, the optimization can be terminated whenever it reaches a feasible result.


\section{Conclusion}

In this work, we presented a complete method for the automated configuration of APS processes. Our approach is based on Gaussian process models which include coarse physical knowledge about the considered application. We validate our models in experiments and observe a predictive performance that is only limited by the process and measurement noise. Our parallelized constrained Bayesian optimization algorithm efficiently directs the search for input parameters that produce the desired coating specifications and maximize the equipment lifetime. The algorithm is based on an acquisition method that has been tailored to the practical needs of APS processes, prioritizing the feasibility of the desired quality parameters. The optimization procedure has been tested on synthetic data and experimentally. It quickly finds feasible combinations and then exploits the newly collected information and the problem structure to minimize the cost. The inclusion of voltage, a status-related measurement, contributes to reducing the harmful effect that components wear and parametric drift have on the process reproducibility. Our method thus anticipates and prevents undesired changes in the properties of the coating. Our results show that carefully designed Bayesian optimization methods are compelling and efficient in the industrial context, especially when the number of trials is severely limited, as in thermal spraying processes. 




\bibliography{references}
\bibliographystyle{ieeetr}

\end{document}